\documentclass[aps,prl,twocolumn,groupedaddress, showpacs]{revtex4-1}
\usepackage{graphicx}
\begin{document}

\title{Picosecond Dynamics of Internal Exciton Transitions in CdSe Nanorods}
\author{D. G. Cooke}
\email{cooke@physics.mcgill.ca}
\affiliation{Department of Physics, McGill University, Montreal, Quebec, Canada H3A 2T8}
\author{P. Uhd Jepsen}
\affiliation{Department of Photonics Engineering, Technical University of Denmark, 2800
Kgs. Lyngby, Denmark}
\author{Jun Yan Lek}
\affiliation{School of Materials Science and Engineering, Nanyang Technological
University, Singapore 639798}
\author{Yeng Ming Lam}
\affiliation{School of Materials Science and Engineering, Nanyang Technological
University, Singapore 639798}
\affiliation{Institute of Materials for Electronic Engineering II, RWTH-Aachen,
Sommerfeldstr. 24, D-52074 Aachen, Germany}
\author{F. Sy and M. M. Dignam}
\affiliation{Dept. of Physics, Engineering Physics and Astronomy, Queen's University,
Kingston, Ontario, Canada K7L 3N6}
\date{\today}

\begin{abstract}
The picosecond dynamics of excitons in colloidal CdSe nanorods are directly
measured via their 1s to 2p-like internal transitions by ultra-broadband
terahertz spectroscopy. Broadened absorption peaks from both the
longitudinal and transverse states are observed at 8.5 and 11 THz,
respectively. The onset of exciton-LO phonon coupling appears as a bleach
in the optical conductivity spectra at the LO phonon energy for times $> 1$
ps after excitation. Simulations show a suppressed exciton temperature due
to thermally excited hole states being rapidly captured onto ligands or
unpassivated surface states. The relaxation kinetics are manipulated and the
longitudinal transition is quenched by surface ligand exchange with hole
capturing pyridine.
\end{abstract}

\maketitle










Quantum dots are one of the essential building blocks of nanoscience and
hold great promise for future optoelectronic applications \cite{Talapin2010}. Their optical properties are conveniently tailored by quantum confinement
of the exciton \cite{Efros2000}. The dynamics of exciton relaxation in
colloidal quantum dots often defines their limitations for a variety of
solution-processed optoelectronic devices, ranging from lasers to spin-cast
solar cells and so is currently a topic of intense research \cite%
{Kambhampati2011}. Our understanding of these processes relies largely on
the interrogation of interband transitions in the optical regime, which
predominantly has been understood in the independent electron and hole
picture. This approximation is valid for particle sizes less than the
exciton Bohr radius where the particle confinement energies well exceed the
Coulomb attraction of the electron and hole. For larger or strongly
asymmetric particles, such as nanorods, this approach is no longer valid and
the interpretation of these spectra becomes more difficult.

Time-resolved THz spectroscopy has made it possible to directly probe
excitons in bulk semiconductors and 2D quantum wells through their internal
transitions on a 1-10 meV scale \cite{Kaindl2003, Suzuki2009, Suzuki2012}.
Excitons in colloidal quantum dot and rod systems, however, are expected to have internal transitions in the 10-50 meV range. THz spectroscopy in the 0.8 - 12 meV band
has been previously used to probe the off-resonant polarizability,
\cite{Beard2002, Wang2006}, long range
coupling \cite{Beard2003}, and energy transfer between electrons and holes in quantum dots \cite{Hendry2006}. Mid-infrared femtosecond spectroscopy has been applied to
probe internal electronic transitions in strongly confined CdSe quantum dots
where the Coulomb interaction with the hole is a small perturbation and the
electronic transitions are well defined on a few 100 meV scale \cite%
{Guyot-Sionnest1998}. These photon energies, however, are too large to probe
the ground state transitions of the exciton in larger particles such as
nanorods.

In this Letter, we apply time-resolved multi-THz spectroscopy in the 1-13 THz (1 - 50
meV) range to probe exciton relaxation dynamics in CdSe colloidal nanorods
capped with organic ligands by probing their internal excitonic transitions.
We observe for the first time intraexcitonic transitions in nanorods and we
see evidence of rapid phonon-induced thermalization in a subset of the rod
population. Two competing channels for exciton relaxation are
observed: capture to surface states on a 4 ps time scale which can be
manipulated through ligand exchange and recombination occurring on a longer
time scale of several 10's of ps. This work demonstrates that multi-THz spectroscopy is
a powerful tool to investigate exciton dynamics in colloidal nanoparticles
and paves the way for future studies of state-dependent relaxation pathways 
\cite{Wong2008}, electron-phonon coupling \cite{Sagar2008} and even
manipulation of excitons in nanoparticles by strong THz fields \cite{ZaksNature2012}.

The samples studied are CdSe nanorods 7 $\pm 1$ nm in diameter and 70 $\pm 10$ nm in length, synthesized using the hot coordinating solvents
method described elsewhere \cite{Lek2011}. The nanorods exhibit an
absorption edge at approximately 685 nm and a weak photoluminesence peak at
683 nm with a FWHM of 30 nm. The surfaces of the nanorods are passivated
with the organic ligand dodecylphosphonic acid (DDPA), which at $\sim 2$ nm in length serves to electrically isolate the nanorods \cite{Beard2003}. These
particles were subsequently drop cast from solution onto indium tin oxide
(ITO) coated glass slides with a sheet resistivity of 2 $\Omega$/sq, which
reflects more than $98\%$ of the THz field amplitude over the entire THz
pulse bandwidth. The thickness of the film varied between 1-1.6 $\mu$m.

\begin{figure}[t]
\includegraphics[width=1.0\columnwidth]{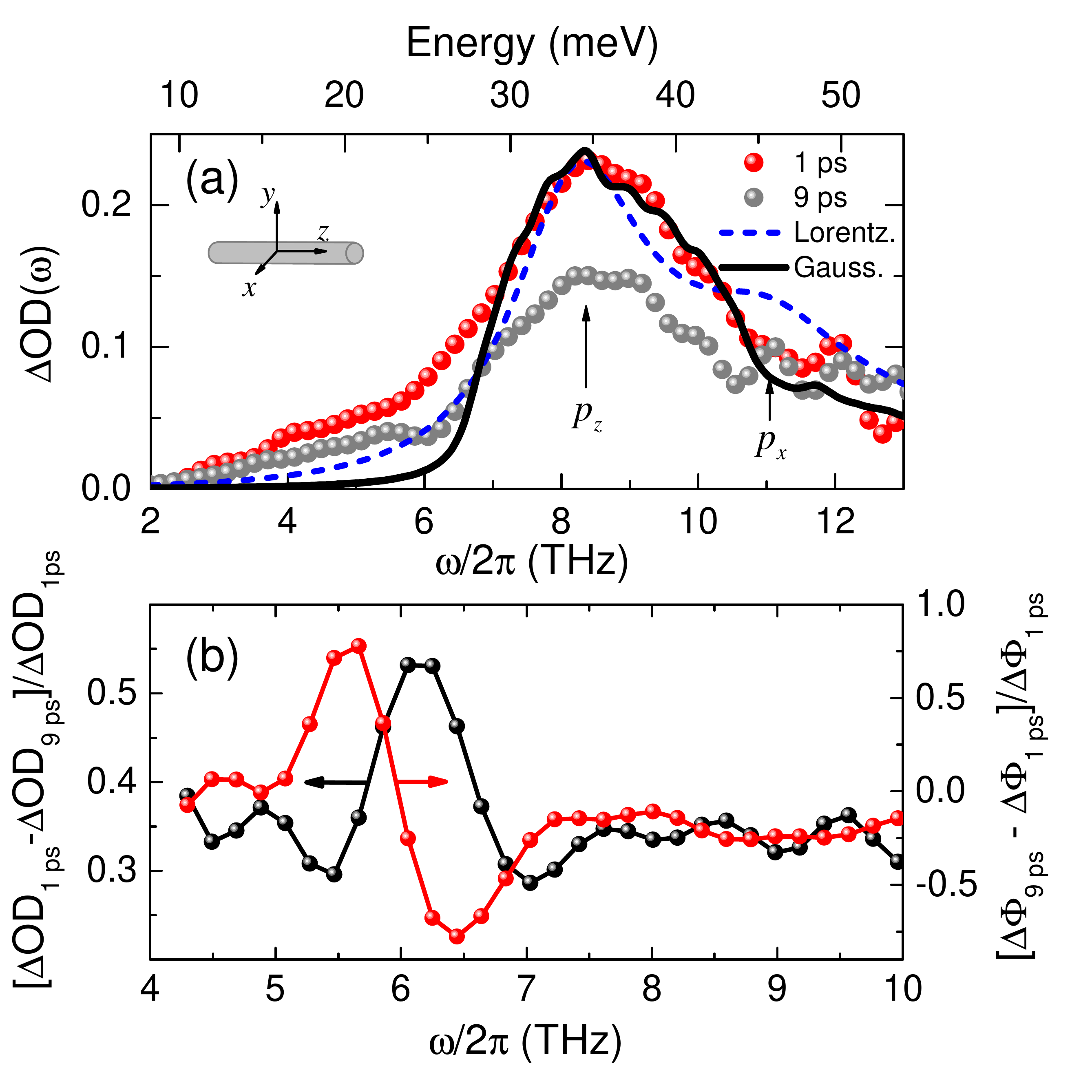}
\caption{(a) Pump induced change in optical density, $\Delta $OD($\protect%
\omega $), of the CdSe nanorod film following 400 nm fs excitation at a
pump-probe delay of $\Delta t=1$ ps and 9 ps. Also shown are calculated
absorbance spectra for nanorods with a radius of 3.5 nm assuming a Lorentzian linewidth (blue dashed) and
for a Gaussian distribution of nanorod radii (black solid) as described in
the text. (b) The relative differential absorbance and phase between 9 ps
and 1 ps pump-probe time delays, showing a bleaching signal at 6.2 THz.}
\label{Fig1}
\end{figure}

The experimental setup is described in more detail elsewhere \cite{Cooke2012}%
, but briefly, we employ an ultra-broadband time-resolved THz spectrometer
operating in reflection mode based on two-color laser plasma THz generation 
\cite{Kress2004} and air-biased coherent detection (ABCD) \cite{Dai2006}.
The spectrometer is driven by 35 fs, 800 nm pulses provided at 1 kHz
repetition rate by a Ti:sapphire regenerative amplifier. The sample was
placed at normal incidence in the focus of the THz pulse and the reflection
was directed via a high resistivity Si beamsplitter to the ABCD module for
detection. A double modulation scheme is used with two lock-in amplifiers to
simultaneously detect the THz transient reflected in the presence of the
pump $E_{pump}(t,\Delta t )$ and the differential transient $\Delta
E(t,\Delta t )=E_{pump}(t,\Delta t )-E_{ref}(t)$ where $E_{ref}(t)$ is the
reflected THz transient in the absence of the photoexcitation and $\Delta t$ is the pump-probe delay time \cite{Iwaszczuk2009}. The nanorods are photoexcited
co-linearly through a small hole in the last parabolic mirror before the
sample by 400 nm fs laser pulses at a fluence of 570 $\mu $J/cm$^{2}$,
unless otherwise stated. All measurements are performed at room temperature
in a dry nitrogen environment.

The differential reflectance can be processed into a frequency dependent
optical density $\Delta $OD$(\omega )=-ln[|E_{pump}(\omega
)|/|E_{ref}(\omega )|]$, with the THz pulse traversing the thin film sample
twice due to the reflection off the underlying ITO which enhances
sensitivity \cite{Gallot2000}. Fig. \ref{Fig1}(a) shows $\Delta $OD$(\omega )
$ at 1 ps and 9 ps following photoexcitation. A broad absorption peak is
observed at 8.5 THz with a higher energy shoulder at approximately 11 THz.
Also shown is the calculated intra-excitonic optical density spectrum (see
below) scaled to fit the 1 ps data; the agreement is very good.

In all of the experimental spectra there is a small but noticeable reduction
in the $\Delta $OD$(\omega )$ spectra at approximately 6.2 THz that only
appears for pump probe times greater than 1 ps (see Fig. \ref{Fig1}(a)) and
that does not appear in the theoretical excitonic absorption results of Fig.
2(a). In Fig. \ref{Fig1}(b) we show the normalized difference between $%
\Delta t=9$ ps and $\Delta t=1$ ps absorbance spectra shown in Fig. \ref%
{Fig1}(a), as well as the differential between the respective phase
functions $\Delta \Phi (\omega )=arg(E_{pump}(\omega ))-arg(E_{ref}(\omega ))
$. \ A clear bleach peak is observed in the absorbance, accompanied by a
phase flip at the CdSe longitudinal optical (LO) phonon energy at 205 cm$%
^{-1}$ (6.15 THz). This bleach signal can be explained if we assume that some
of the of nanorods have an intra-excitonic transition energy (perhaps $%
1s-2p_{z}$) that is resonant with the LO phonon. In these nanorods, there
will be a resonant Fr\"{o}hlich interaction scattering excitons between the
two transition states via the absorption of an LO phonon \cite%
{Mittleman1994, Wise2000, Sagar2008}. This process will lead to a rapid
thermal population of the upper state, which will result in the observed
bleaching of the absorption at $6.2$ THz. The linewidth of the bleach signal
is $\sim $0.5 THz, consistent with previous measurements of the Raman
linewidth in CdSe nanorods \cite{Lange2008}. The onset of exciton-phonon coupling has been
previously observed and is attributed to rapid trapping of charge on the
surface of the nanorod on a 1-10 ps time scale, causing an internal field
which polarizes the exciton and leads to an increase in exciton-phonon
coupling \cite{Wise2000, Sagar2008}. For nanorods, however, such an internal
field is not necessary to initiate the exciton-phonon coupling due to the
asymmetric wavefunctions. Thus we attribute the
onset to the thermalization of the excitons due to their $\sim 1.3$ eV
excess energy compared to the ground state.

To calculate the CdSe nanorod absorbance spectrum arising from transitions
between excitonic states, we first calculate the light and
heavy hole excitonic energies and envelope functions by expanding them
in a basis of non-interacting electron and hole nanorod states. \ We neglect
the coupling between light holes (LH) and heavy holes (HH) and calculate the electron-hole
Coulomb interaction matrix elements to obtain the ground $1s$ excitonic
state as well as all of the excited states up to energies beyond the
frequency response of the THz detector. In contrast to small quantum dots,
the electron and hole states are greatly modified by the Coulomb interaction
between them and we require a basis of 10 electron, 60 LH and
60 HH states to obtain convergence for the relevant excitonic
states. \ We employ infinite confining potentials at
the surfaces of the cylindrical nanorods and use effective masses given by
the diagonal terms in the Luttinger matrix of Ref. \cite{Ekimov1993}. \ We
take the nanorods to be cylindrical with a length of 70 nm and assume that
there is at most one exciton per nanorod. \ It is assumed the nanorods all have
their cylindrical axes parallel to the substrate surface; for a give
nanorod, the direction parallel to the cylindrical axis is the $z$-direction and
the radial direction in the plane of the substrate is the $x$%
-direction (see inset to Fig. \ref{Fig1}(a)). \ Because the LHs are heavier in the
radial direction, the lowest-energy exciton is the $1s$ LH exciton since
the ground state HH exciton is 16.1 meV higher in energy.

\begin{figure}[t]
\includegraphics[width=1.0\columnwidth]{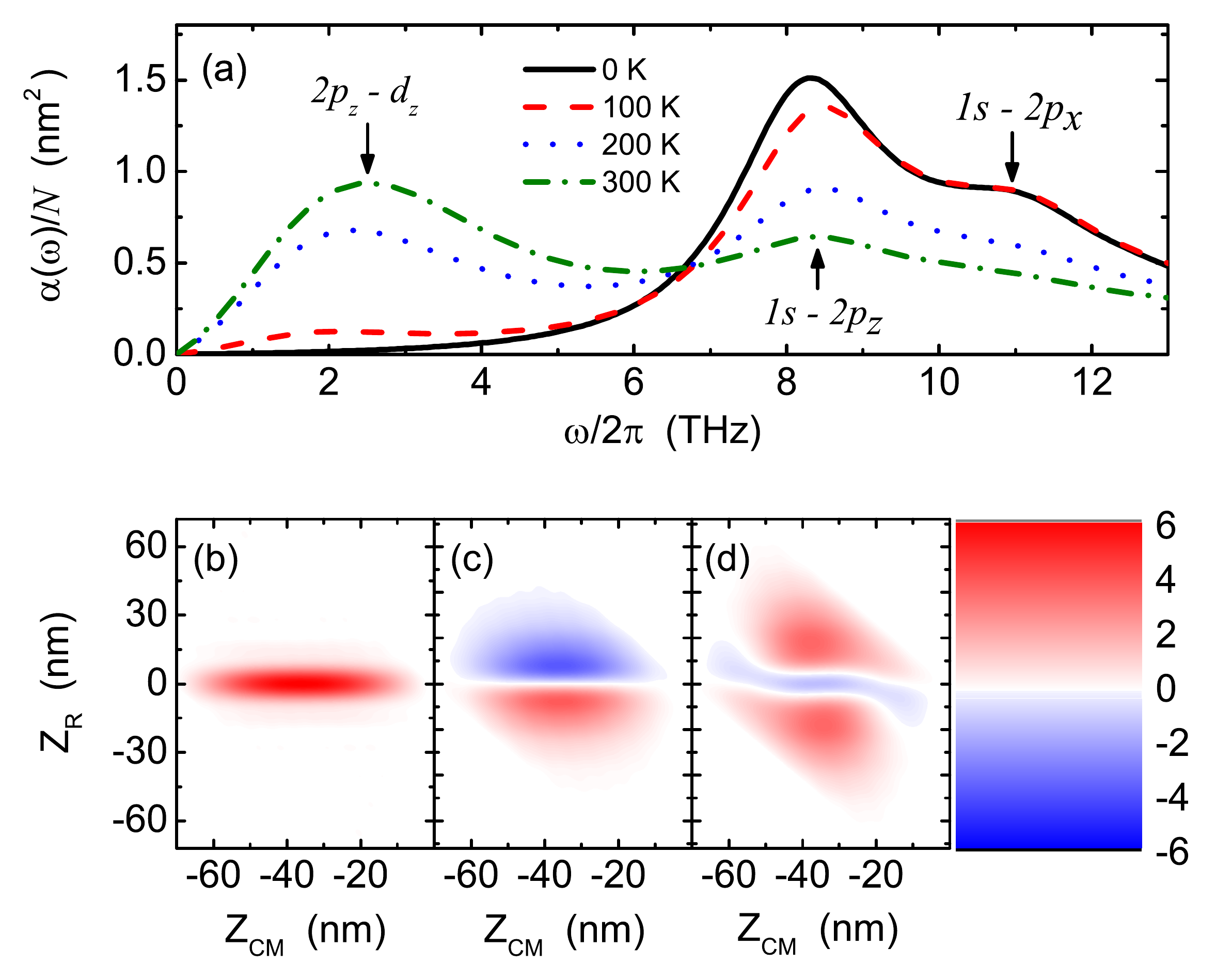}
\caption{{}(a) The calculated absorbance spectra for four different
temperatures for nanorods with a radius of 3.5 nm, assuming a Lorentzian
linewidth of 10 meV. The transition for the three main peaks indicated.
The exciton wavefunction is plotted as a function of the
center of mass ($Z_{CM}$) and relative ($Z_{R}$) for the (a) $1s$, (b) $%
2p_{z}$ and (c) $d_{z}$ excitonic states.}
\label{Fig2}
\end{figure}

The dashed blue curve in Fig. 1(a) is produced assuming a rod radius of 3.5
nm and a Lorentzian linewidth of 2.4 THz (10 meV) for all transitions and
scaled to fit the main peak for the 1 ps data. \ To produce this spectrum,
we assume that all excitons are in the $1s$ LH ground state when the THz
pulse arrives. The main absorption peak at 8.3 THz is due to the transition
from the $1s$ LH exciton ground state to a $2p_{z}$-like excitonic state. \
The $z$-dependence of these two states on the rod axis is shown in Figs.
2 (b) and (c) respectively. The dipole matrix element for this transition is 
$D_{z}/e=3.0$ nm, where $e$ is the magnitude of the electron charge. \ The
weaker transition, which appears as a shoulder at 10.8 THz is due to a
transition from the LH $1s$ ground state to the LH $2p_{x}$ state, which has
a dipole moment of only $D_{x}/e=1.15$ nm. \ We note that there
are many other excitonic states with excited center-of-mass motion with
lower energies than the $2p_{z}$ and $2p_{x}$ states, but the transition
dipole between these states and the ground state is negligible.

If we now assume that the excitons are in quasi-thermal equilibrium, we
obtain the absorbance spectra shown in Fig. 2(a) for four different
temperatures. As the temperature increases,  the relative strengths of the $%
1s-2p_{z}$ and $1s-2p_{x}$ transitions change. In addition, above about $%
T=100$ K, a new peak appears at about $2.4$ THz, which is largely due to a
transition from the LH $2p_{z}$ state to a higher LH $d_{z}$-like state
shown in Fig. 2(d); the dipole matrix element for this transition is $%
D_{pd}/e=12.0$ nm, which is why the peak is so large. 

Reasonable agreement with experiment is only found if we
assume an effective temperature of less than 100 K. \ We interpret this
result as a strong indication that only the ground state exciton has a
significant population when the THz pulse hits the sample. \ We propose that
the absence of excitons in excited states is due to the rapid transfer of
the holes in excited excitonic states into traps in the ligand. \ As such
trapped holes are spatially separate from the electrons in the nanorods,
these "bound excitons" have negligible intra-excitonic dipole matrix
elements and so do not contribute to the THz absorption. 

Scanning electron microscopy (SEM) of the nanorod samples shows a
distribution of nanorod diameters and length. \ For the range of lengths
observed, the THz absorption shows little length-dependence but significant dependence on radius. The
solid black curve in Fig. 1(a) is produced taking the nanorod radii 
to have a Gaussian distribution, with a mean of 3.25 nm and standard
deviation of 0.7 nm, consistent with the values observed in SEM. \
To make the calculations tractable, we have discretized the distribution in
steps of 0.05 nm, which results in   small oscillations seen in the curve.
We have measured the thickness of the nanorod film in the measured spot to be $\sim1.5$ $\mu $m. The
remaining unknown is the exciton density, treated as a fitting
parameter and corresponding to 0.45
excitons per nanorod for the $\Delta t = 1$ ps spectra. \ The agreement is quite
good, except for frequencies below $\sim 6$ THz where a low frequency absorption tail is present. This may be due to contributions from excited state transitions or a non-Gaussian, large-radius tail in the actual nanorod radius distribution.

\begin{figure}[t!]
\includegraphics[width=\columnwidth]{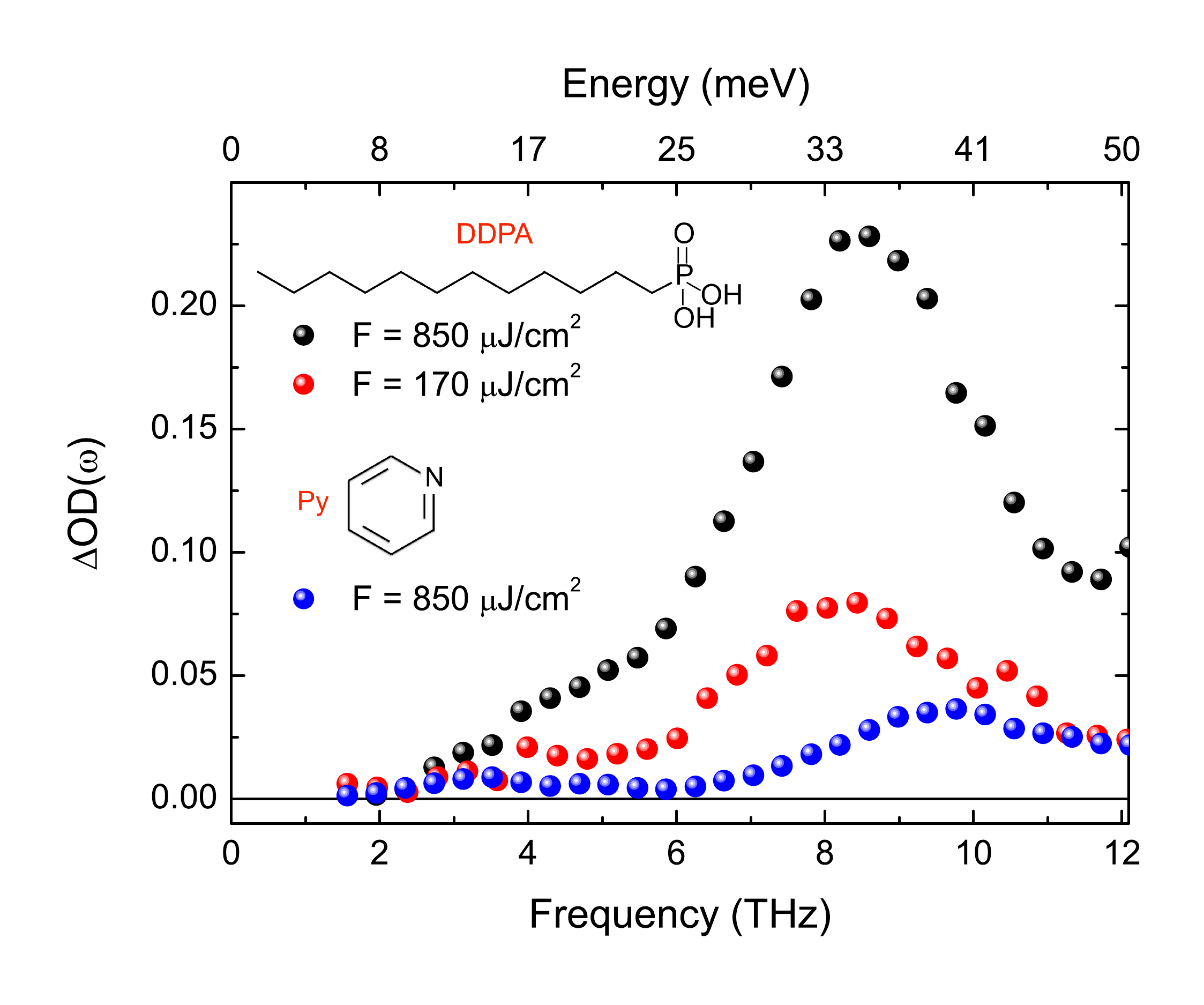}
\caption{$\Delta$OD($\omega$) spectra for the DDPA passivated rods, as well as the effect of partial ligand exchange of DDPA with pyridine. Pump fluences are indicated in the figure.}
\label{Fig3}
\end{figure}

To investigate the influence of surface passivation, we have performed a
ligand exchange with a hole scavenging ligand following the work of
Guyot-Sionnest et al., where 70$\%$ of the DDPA is replaced by pyridine \cite{Guyot-Sionnest1998}.
Fig. \ref{Fig3} shows there is a strong suppression of the excitonic $\Delta$OD peak
and a slight blue shift to 9.8 THz. At this point there are no models for
the influence of bound surface charge on the internal excitonic transitions
of a nanorod. It appears that these states have lower energy
than the excited excitonic states of the nanorods, as we see from comparison
of the room temperature plot in Fig. \ref{Fig2}(a) to the experimental results that
there is essentially no evidence of populations in the excited excitonic
states at room temperature. Also shown in Fig. \ref{Fig3} is the $\Delta$OD($\omega$) spectra of the DDPA passivated nanorods at $\Delta t = 1.0$ ps for a factor of 5 reduction in pump fluence. No shift in the peak frequency is observed with reduced excitation density, inconsistent with a plasmon - Fr\"ohlich phonon mode \cite{Hyun2011}, and consistent with our assumption of there being no more than one exciton per nanorod in our model.

\begin{figure}[t!]
\includegraphics[width=0.9\columnwidth]{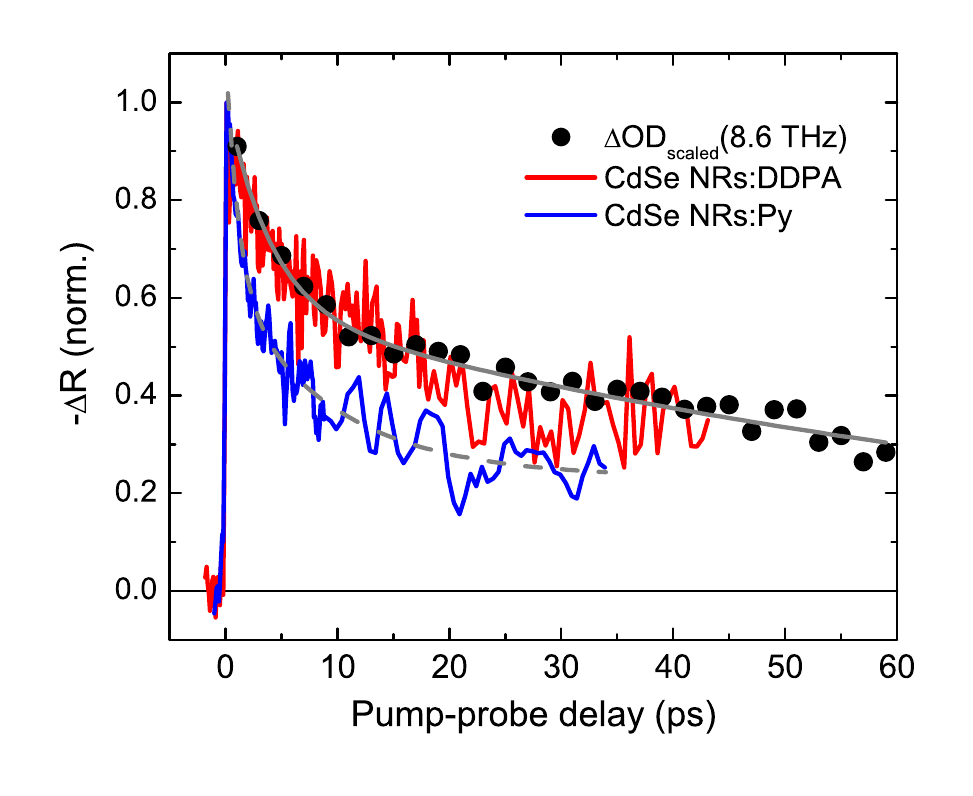}
\caption{The normalized and scaled $\Delta$OD(8.5 THz) and the differential THz reflectance ($\Delta R(t)$) dynamics for the DDPA and Pyridine passivated CdSe nanorods.}
\label{Fig4}
\end{figure}

To follow the exciton relaxation dynamics, a one dimensional slice at the peak (8.6
THz) of the $\Delta$OD spectra is shown in Fig. \ref{Fig4}, demonstrating
a clear bi-exponential decay with 42 $\%$ of the signal undergoing a fast
4.0(7) ps decay and the remaining 58 $\%$ relaxing slower with a 90(8) ps
decay constant. These relaxation times are comparable with previous
time-resolved optical studies of CdSe nanorods \cite{Mohamed2001, Robel2006}. The change in exciton relaxation dynamics due to ligand exchange was also
investigated by performing a less time-consuming differential reflectivity
measurement, monitoring the peak of the differential THz waveform shown in
Fig. \ref{Fig1} for the DDPA and pyridine ligand passivated nanorods. The
normalized transients for both samples are shown in Fig. \ref{Fig4}, which track the $\Delta $R($\Delta t $) transient for the CdSe:DDPA nanorod and confirms that 
$\Delta $R($\Delta t $) is proportional to the ground state
population. Ligand exchange with pyridine noticeably decreases the lifetime
of the exciton ground state, well described again by a bi-exponential decay
with a 1.8(2) ps fast time constant and a slower 42(3) ps time constant not well resolved by our limited temporal range. The fast decay is therefore identified with rapid hole charge transfer and localization from the core to the pyridine molecule, forming a charge separated complex and leaving a core electron behind.

In conclusion, we have determined the picosecond dynamics of excitons in CdSe nanorods by probing their internal excitonic transitions at meV energies. The exciton ground state decays on two time scales, one dominated by surface trapping on a sub-5 ps
time scale and another on a several 10's of ps time scale. The dynamics of
these processes are significantly influenced by the capping by
organic ligands, with the transition strength suppressed and relaxation
enhanced by introduction of the hole scavenging pyridine ligand. Several
promising lines of investigation are now possible using these internal
exciton transitions. Signatures of electron-phonon coupling are indicated
when there are absorbance features occurring at the nanorod LO phonon
frequency. The influence of surface ligands on dynamics as well as the
energetics of the excitonic transitions will lead to a sensitive test of
theory. Finally, as the temporal resolution is improved it will be possible to investigate directly the cooling of hot excitons through the densely packed manifold of states.

\begin{acknowledgments}
M. M. Dignam and D. G. Cooke acknowledge funding from NSERC. D. G. Cooke
also acknowledges funding from CFI and FRQNT. The authors thank M. Kraus and
P. Kambhampati for photoluminescence measurements of the nanorods.
\end{acknowledgments}



%

\end{document}